\documentclass[reprint,aps,nofootinbib,superscriptaddress,prb]{revtex4-2} 
\usepackage{natbib}
\usepackage[utf8]{inputenc}
\usepackage{amsmath,amssymb}
\usepackage{mathtools} % Provides the very useful \mathclap
\usepackage{mathrsfs} %fancy C

\usepackage{xcolor}
\usepackage[]{colorspace}
\usepackage{graphicx}
\usepackage{hyperref}
\usepackage{microtype}
\usepackage{csquotes} 
\usepackage[newzealand]{babel}
\usepackage{lmodern} 
\usepackage[T1]{fontenc} 
\usepackage[locale = UK]{siunitx}
\newcommand{\defi}{\mathrel{\mathop:}=} % LHS defined by RHS
\newcommand{\tb}{\textbf}
\newcommand{\mb}{\mathbf} % For various reasons it's better to keep the two separate; especially once one wants to move things into beamer slides, as there textbf will produce sans serif fonts
\newcommand{\Tr}{\ensuremath{\operatorname{Tr}}} % The trace command
\newcommand{\bPsi}{\boldsymbol{\Psi}} % Self-explanatory unless we change notation

\newcommand{\bPhi}{\boldsymbol{\Phi}} % Self-explanatory unless we change notation
\newcommand{\WbPsi}{\widetilde{\boldsymbol{\Psi}}}
\newcommand{\WbPhi}{\widetilde{\boldsymbol{\Phi}}}
\newcommand{\Wa}{\widetilde{a}}

\newcommand{\brr}[1]{\left( #1 \right)} % Round brackets
\newcommand{\helB}{Z}
\newcommand{\dif}{\mathop{}\!\mathrm{d}}

\newcommand{\ldoc}{\mathscr{C}}
\newcommand{\ldocbar}{\mathscr{C}}
\newcommand{\ldocoa}{\langle\mathscr{C}\rangle}
\newcommand{\ldocoabar}{\langle\overline{\mathscr{C}}\rangle}

\newcommand{\tmatrix}{\emph{T}-matrix}
\definecolor{darkgreen}{rgb}{.125,.5,.25}

\begin{document}
%______________The macros of RevTeX 4.2 for author, affiliation, title, abstract, etcpp follow
% \captionsetup[figure]{labelfont={bf},labelformat={default},labelsep=period,name={FIG.}}
\title{Orientation Averaging of Optical Chirality Near Nanoparticles and Aggregates} 

\author{Atefeh \surname{Fazel-Najafabadi}}
\email{atefeh.fazelnajafabadi@vuw.ac.nz}
\affiliation{School of Chemical and
  Physical Sciences, Victoria University of Wellington, PO Box 600,
  Wellington 6140, New Zealand}
\affiliation{The MacDiarmid Institute for
  Advanced Materials and Nanotechnology}
\author{Sebastian Schuster}
\email{sebastian.schuster@sms.vuw.ac.nz}
\affiliation{School of Chemical and
  Physical Sciences, Victoria University of Wellington, PO Box 600,
  Wellington 6140, New Zealand}
\affiliation{The MacDiarmid Institute for
  Advanced Materials and Nanotechnology}
\affiliation{School of Mathematics and Statistics, Victoria University of Wellington, PO Box 600, Wellington 6140, New Zealand}
\author{Baptiste Auguié}
\email{baptiste.auguie@vuw.ac.nz}
\affiliation{School of Chemical and
  Physical Sciences, Victoria University of Wellington, PO Box 600,
  Wellington 6140, New Zealand}
\affiliation{The MacDiarmid Institute for
  Advanced Materials and Nanotechnology}
\date{\today}
\begin{abstract}
% \begin{itemize}\it
% \item Chirality and nano-optics
% \item Optimisation would be great, but difficult
% \item The \tmatrix\ method, but no result yet
% \item We derive results, test, and discuss
% \item Open new possibilities
% \end{itemize}
Artificial nanostructures enable fine control of electromagnetic fields at the nanoscale, a possibility that has recently been extended to the interaction between polarised light and chiral matter. The theoretical description of such interactions, and its application to the design of optimised structures for chiroptical spectroscopies, brings new challenges to the common set of tools used in nano-optics. In particular, chiroptical effects often depend crucially on the relative orientation of the scatterer and the incident light, but many experiments are performed with randomly-oriented scatterers, dispersed in a solution.
We derive new expressions for the orientation-averaged local degree of optical chirality of the electromagnetic field in the presence of a nanoparticle aggregate. This is achieved using the superposition \tmatrix\ framework, ideally suited for the derivation of efficient orientation-averaging formulas in light scattering problems. Our results are applied to a few model examples, and illustrate several non-intuitive aspects in the distribution of orientation-averaged degree of chirality around nanostructures. The results will be of significant interest for the study of  nanoparticle assemblies designed to enhance chiroptical spectroscopies, and where the numerically-efficient computation of the averaged degree of optical chirality enables a more comprehensive exploration of the many possible nanostructures.
\end{abstract}
\maketitle
\section{Introduction}
%
% \begin{itemize}\it
% \item Why chirality and plasmonics are seeing such a booming interest. 
% \item - Progress in nanofabrication, understanding of the origins of chirality, prospects of enhanced spectroscopies in a variety of systems, fabricated by top down or bottom up approaches.
% \item What are the bottlenecks in this area? Where can modelling help?
% \item The \tmatrix\ method, against alternatives
% \item What we do here: present, derive, test, explain, discuss
% \item Outlook: what becomes possible, what questions can be addressed
% \end{itemize}

% \printinunitsof{in}\prntlen{\columnwidth}\\ %3.4
% \printinunitsof{in}\prntlen{\textwidth} %7

Chiral materials naturally respond with a slight asymmetry to left or
right circularly-polarised light~\cite{Barron:2009aa,Berova:2000aa}, and the minute difference in spectroscopic signals -- in absorption, fluorescence, Raman scattering etc. -- can be exploited to trace back precious 3D structural information at the nanoscale, such as the relative orientation of molecular groups (handedness). This is particularly important in the realm of biochemistry~\cite{ProteinGreenfield2006using,DrugNguyen2006chiral}, where mirror-image molecules (enantiomers) can differ dramatically in their interaction with other chiral molecules. The ability to characterise the handedness of chiral molecules with optical spectroscopy, in the form of circular dichroism or related techniques, remains a challenging pursuit with important practical applications~\cite{Berova:2012aa}. The main limitation is that the chiroptical activity of most molecules is very weak; most naturally-occurring chiroptical effects have thus been restricted in their study and application to bulk samples such as highly-concentrated chemical solutions. 

Artificial chiral nanostructures have opened remarkable new perspectives in this area~\cite{Amabilino:2009aa}, with demonstrations of enhanced chiroptical signals~\cite{Hentschele1602735,Mun:2020aa}. The rising interest in nano-optics in the past few decades has highlighted the potential of nano-materials and nano-structures engineered to bridge the gap between the wavelength of free propagating light and localised optical near-fields exciting molecules. Acting as nano-antennas, metallic or dielectric nanoparticles supporting resonant electromagnetic modes can funnel light into subwavelength regions associated with greatly amplified electromagnetic fields~\cite{Novotny:2006aa}. This effect has been a cornerstone in the development of surface-enhanced Raman (SERS) or fluorescence spectroscopies~\cite{LeRuE08}. 
Such nanostructures can similarly enhance the chiroptical response of a chiral molecule~\cite{schaaff2000giant,Lieberman:2008p113,govorov2010theory,govorov2011plasmon,PhysRevB.87.075410,SiNanodiskDimerZhao2019,najafabadi2017analytical}

Cohen and coworkers~\cite{Yang:2009p286,TangCohen2010} recognised that engineered nanostructures may enhance the interaction between light and chiral matter not just through a locally-enhanced magnitude of electric ($\mb{E}$) and magnetic ($\mb{B}$) fields, but also conjointly by shaping the local degree of optical chirality $\ldoc\propto\Im(\mb{E}^*\cdot\mb{B})$, where $\Im$ stands for \enquote{imaginary part} --- a quantity requiring full consideration of the vectorial nature of both fields and their complex phase~\cite{SchaeferlingTailoringChirality}. 

Specifically, let us consider the time-averaged rate of absorption $A$ of a chiral molecule excited by local electric $\mb{E}$ and magnetic fields $\mb{B}$ at a frequency $\omega$, and characterised by its electric dipole moment $\mathbf{p}=\alpha\mb{E}-i\gamma\mb{B}$, and magnetic dipole moment $\mb{m}=\beta\mb{B}+i\gamma\mb{E}$~\cite{Hentschel2013Complex2},
\begin{equation}
	A \defi \frac{\omega}{2}\Im\left({\mb{E}}^*\cdot {\mb{p}} + {\mb{B}}^*\cdot {\mb{m}}\right),
\end{equation}
where $\alpha$ is the electric dipole polarisability, $\beta$ the magnetic dipole polarisability, and $\gamma$ characterises the mixed electric-magnetic dipole polarisability and is responsible for optical activity. Under the common simplifying assumption of monochromatic fields and a homogeneous embedding medium, this expression simplifies to~\cite{Hentschel2013Complex2}:
\begin{equation}
	A =  \frac{\omega}{2} \left[ \Im\left(\alpha\right) |{\mb{E}}|^2 + \Im\left(\beta\right) |{\mb{B}}|^2 + \Im\left(\gamma\right) \ldoc\right],
\end{equation}
where %
\begin{equation}
\ldoc = \frac{-\omega\epsilon_0}{2}\Im(\mb{E}^*\cdot\mb{B}) \label{eq:OC1}
\end{equation}
is referred to as the local degree of optical chirality (LDOC), first introduced by Lipkin~\cite{LipkinZilch}. 

While $\alpha,\beta,\gamma$ are intrinsic properties of the scattering particle or molecule, $\ldoc$ encapsulates the chirality of the electromagnetic field at a given position in space~\cite{LipkinZilch}. The differential absorbance of a molecule in response to a plane wave with left and right circular polarisations will be proportional to $\Im\left(\gamma\right) \ldoc$, where $\ldoc=\pm \frac{k\varepsilon_0E_0}{2}$, with $k$ the wavenumber, $\varepsilon_0$ the permittivity of free space, and $E_0$ the amplitude of the incident electric field. By placing the molecule in the vicinity of a nanostructure, the local electromagnetic field, and therefore $\ldoc$, can be amplified beyond this value~\cite{TangCohen2010,SchaeferlingTailoringChirality}. This offers the possibility to engineer nanostructures that optimise the chiroptical response of analytes. Such designs can be modelled by solving Maxwell's equations for a given nanostructure, calculating $\mb{E}$ and $\mb{B}$ which readily provide $\ldoc$; this procedure has been used to estimate the enhancement of chiroptical signals for molecules in a variety of configurations~\cite{Hendry:2010aa,Hentschel2013Complex2}.

% The deliberate design of nanostructures to maximise the local degree of chirality thus introduces a new degree of freedom benefiting all chiroptical spectroscopy techniques. Such artificial structures are also expected to yield entirely new responses, not present in natural systems\cite{}. These intriguing predictions are not yet fully understood and better theoretical tools are needed\cite{}, as well as the exploration of a vast range of possible structures\cite{}.

Although the design of useful structures benefits from extensive research and advances in surface-enhanced spectroscopies and nano-optics in the past decades, the consideration of chirality at the nano-scale, and its interaction with polarised light, brings a host of new challenges to our intuition, and our modelling capabilities. Where surface-enhanced spectroscopies such as SERS rely primarily on the amplification of the electric field intensity in the immediate vicinity of nanostructures, in pursuing enhanced optical activity one must also consider the polarisation state of near-field regions, which remains generally less intuitive, and for which we have fewer helpful rule-of-thumbs to guide us in the design of efficient chiroptical platforms~\cite{sym11091113,Lasa-Alonso:2020aa}. 

\begin{figure}
	\centering
\includegraphics[width=\columnwidth]{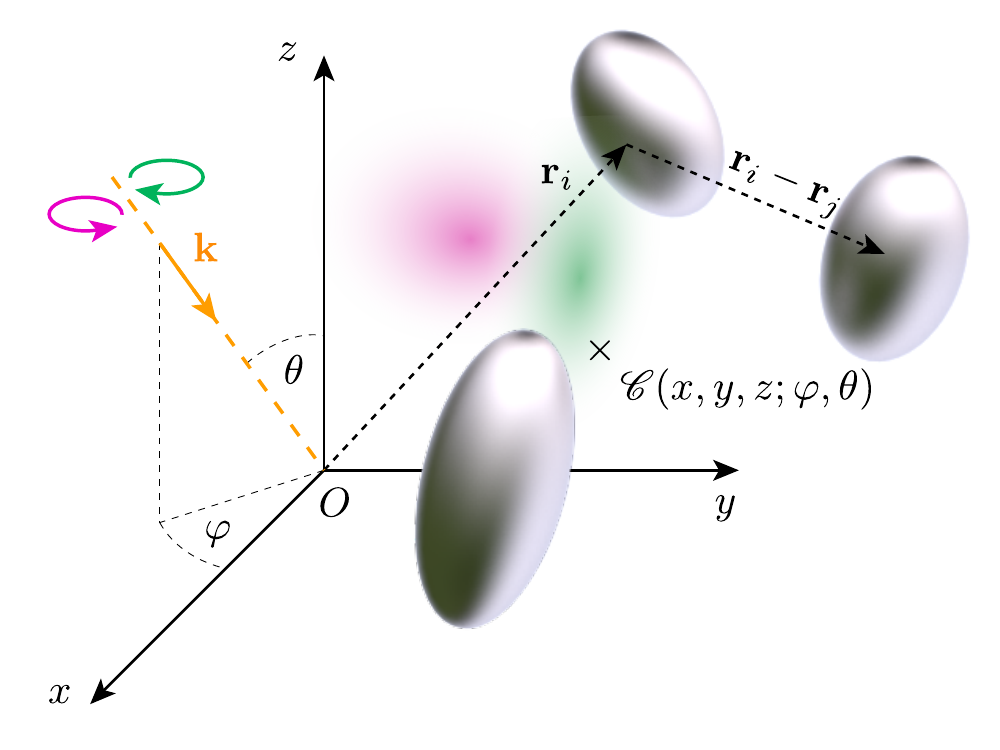}
	\caption{Schematic illustration of the light scattering problem under consideration. A rigid cluster is formed by $N$ particles arranged in a fixed spatial configuration; each particle may be non-spherical and oriented arbitrarily with respect to the common \enquote{cluster} reference frame $(O,x,y,z)$. The orientation and position $\mathbf{r_i}$ of each particle ($i=1\dots N$) is arbitrary but fixed in this reference frame; in practice such particles would be held in place by a template~\cite{Guerrero-Martainez:2011ac,Kuzyk:2018aa,Lan:2018aa}, our calculations assume that the template has no impact on the optical properties of the structure. Incident light takes the form of a plane wave with wavevector $\mathbf{k}$ from an arbitrary direction $(\varphi,\theta)$ in the cluster reference frame, and we consider both possible states of circular polarisation (left, $L$, or right $R$). For each direction of incidence and either polarisation we can calculate the local degree of chirality $\ldoc$ in the vicinity of the particles, and map its spatial distribution (schematically represented by purple and green gradients). The quantity $\langle\ldoc\rangle$ derived in this work represents the average value for the chosen polarisation when the cluster as a whole is randomly oriented with respect to incident light, or equivalently, when $\mathbf{k}(\varphi,\theta)$ spans the full solid angle in the cluster reference frame.}
	\label{fig1}
\end{figure}

An additional challenge presented by chiral nanostructures is the particular importance of the relative orientation between incident light and the scatterer. For example, a 2D structure fabricated by lithography may present a differential response to circularly polarised light known as \enquote{extrinsic} chirality, though the same structure would display no circular dichroism if illuminated from all directions of incidence. This distinction is particularly relevant as many 3D chiral nanostructures are characterised in solution, where they are randomly-oriented with respect to incident light. To describe the optical response of such samples it is essential to be able to reliably compute the orientation-averaged response of the structure to incident light of a given polarisation (Fig.~\ref{fig1}). This can present a time-consuming hurdle for theoretical calculations. In the context of optical activity, a comparison with experiments may involve computing far-field orientation-averaged circular dichroism~\cite{OrientationAveragedMetaMaterials}, and for experiments testing the effect of superchiral fields on molecular species, the orientation-averaged local degree of optical chirality $\langle \ldoc\rangle$; the two quantities are not necessarily directly related to each other~\cite{OCandCDgarcia2019enhanced}.

Many common tools used in nano-optics, such as the Finite-Element Method (FEM)~\cite{volakis1998finite}, Finite-Difference Time Domain (FDTD)~\cite{taflove2005computational}, Discrete-Dipole Approximation (DDA)~\cite{draine1994discrete}, Volume or Surface Integral Equation methods (VIE/SIE)~\cite{kern2009surface,botha2006solving,Reid:2015aa}, solve the Maxwell equations for a given configuration of particles and a given incident field. These computations can be demanding even for just a few particles, and improving the methods is still an active area of research. This computational burden limits the ability to explore the vast landscape of possible structures and seek those that approach optimum bounds~\cite{Kramer:2017aa,Fernandez-Corbaton:2016aa}; it also often leads to shortcuts where accuracy is traded for practical computational time. For example, orientation-averaged properties may be approximated by considering three orthogonal directions of incidence ($x$, $y$, $z$ axes), though the validity of this approximation is not clear and has received little attention to date. A rigorous result for the orientation-averaging of observable quantities such as far-field cross-sections, near-field intensity, or in our case here the LDOC $\ldocoa$ considers the integral over Euler angles $\varphi,\theta$ describing the direction of the wavevector incident on the scatterer (or collection of scatterers) (Fig.~\ref{fig1}),

\begin{equation}
\langle\ldoc(x,y,z)\rangle=\frac{1}{4\pi}\int_0^{\pi} \int_0^{2\pi}  \ldoc(x,y,z;\varphi,\theta)  \sin\theta \dif \varphi \dif \theta .
\label{eq:eulerOA}
\end{equation}

This averaging over the full solid angle of possible illumination directions is often complemented by a further averaging over the direction of polarisation for linearly-polarised light~\cite{mishchenko1989interstellar, mishchenko1990extinction,Khlebtsov92, borghese2007scattering}. In the case of circular polarisation, however, the rotations of the wavevector do not affect the helicity of the light and it is therefore useful to consider the angular averaging over $(\varphi,\theta)$ for left ($L$) or right ($R$) polarisations separately~\cite{Suryadharma18}. The fully averaged optical response to unpolarised light will be the average of both.

In contrast to the methods previously mentioned, the \tmatrix\ framework affords a relatively efficient solution to the Maxwell equations in a basis of vector spherical waves~\cite{MishchenkoTL02}; intrinsic to the method is therefore the description of a scatterer or collection of scatterers, to waves incident from any direction. This is a particular strength of the method which has led to analytical formulas for orientation-averaged quantities such as far-field cross-sections~\cite{Khlebtsov92,Mackowski1994SphereClusters}. Such results were recently extended by Rockstuhl and coworkers for optical activity~\cite{OrientationAveragedMetaMaterials}, and by Stout and co-workers for near-field quantities in clusters of spherical particles~\cite{StoutTransferMatrixMultipleScatterer2002,StoutTransferMatrixMultipleScatterer2008}.
We extend here these previous results to the calculation of orientation-averaged near-field optical chirality. The rest of this manuscript is divided as follows. First, we summarise the essential formulas and our notations for the superposition \tmatrix\ method, followed by our newly-derived formulas. In a second part, we illustrate these results by applying the formulas to model clusters of nanoparticles, and discuss the distribution and enhancement of near-field optical chirality around various nanostructures.
These results illustrate the power of analytical results enabled by the \tmatrix\ method, which can provide rapid, accurate, and physically-insightful simulations of a vast range of nanostructures. We hope to encourage more researchers to adopt this method in their comparisons to experiments, but also to further improve the method beyond its current limitations~\cite{Schebarchov:2019aa,PhysRevA.96.033822}.

% also help understand C, its relation to CD, and optimise structures

%
\section{\texorpdfstring{\tmatrix\ formalism and main formulas}{T-matrix formalism and main formulas}} 
We follow standard notations~\cite{MishchenkoTravisLacisMain,LeRuE08} for vector spherical waves and describe the geometry in the usual spherical coordinates $(r,\varphi,\theta)$ (Fig.~\ref{fig1}); for the extension to $N$ scatterers we follow closely the scatterer-centred superposition \tmatrix\ approach presented by Auger and Stout~\cite{StoutAD08,StoutTransferMatrixMultipleScatterer2008}. This section provides a brief summary of our choice of definitions for completeness.
\subsection{A single scatterer}
Waterman's \tmatrix\ method~\cite{waterman1965matrix, waterman1969new} offers a rigorous solution of Helmholtz' wave equation by describing the light scattering process in terms of an incident field and a scattered field, both expanded into bases of regular and irregular Vector Spherical Wave Functions (VSWFs), respectively~\cite{StoutTransferMatrixMultipleScatterer2002},
% In the \tmatrix\ framework the incident and scattered electric fields are expanded 
% As we deal with linear media, without loss of generality, we can work with monochromatic and time-harmonic fields, allowing us to neglect the time-dependence. In this instance, one can write any solution as a linear combination of vector spherical wave functions as 
%
\begin{alignat}{2}
	\mb{E}_{\text{inc}}(k\mb{r})&=\sum_{n=1}^{\infty} \sum_{m=-n}^n[&&\Wa_{1,nm}\widetilde{\mb{M}}_{nm}\!(k\mb{r})\nonumber\\
	&&&+ \Wa_{2,nm}\widetilde{\mb{N}}_{nm}\!(k\mb{r})] \label{eq:E-field}\\
	\mb{E}_{\text{sca}}(k\mb{r})&=\sum_{n=1}^{\infty} \sum_{m=-n}^n[&&{f}_{1,nm}{\mb{M}}_{nm}(k\mb{r})\nonumber\\
	&&&+ f_{2,nm}{\mb{N}}_{nm}(k\mb{r})] 
\end{alignat}
\noindent
where $\widetilde{\mb{M}}_{nm}$ and $\widetilde{\mb{N}}_{nm}$ are normalised vector spherical wave functions composed of spherical harmonics ($Y_{nm}(\varphi, \theta)$) and spherical Bessel ($j_n(k\mb{r})$ functions for the ingoing wave, while ${\mb{M}}_{nm}$ and ${\mb{N}}_{nm}$ are composed of spherical harmonics and spherical Hankel functions ($h_n(k\mb{r})$) of the first kind for the scattered wave (See Appendix~\eqref{app:A} for their explicit definition, following Stout and coworkers~\cite{StoutTransferMatrixMultipleScatterer2002,StoutAD08,StoutTransferMatrixMultipleScatterer2008}). $\Wa_{1,nm}$, $\Wa_{2,nm}$, $f_{1,nm}$, and $f_{2,nm}$ are the associated expansion coefficients. 

In a more compact form, we write the incident and scattered electric fields as:
\begin{align}
&\mb{E}_{\text{inc}}(k\mb{r})=\widetilde{\bPsi}(k\mb{r})\widetilde{\mb{a}} \label{eq:E-field2}\\
&\mb{E}_{\text{sca}}(k\mb{r})=\bPsi(k\mb{r}) {\mb{f}} \label{eq:Esca}  
\end{align}
where ${\widetilde{\mb{a}}, \mb{f}}$ are (infinite-dimensional) column vectors of incident and scattered field coefficients. $\widetilde{\bPsi}=[\widetilde{\mb{M}} \quad \widetilde{\mb{N}}], \bPsi = [\mb{M} \quad \mb{N}]$ are the corresponding, infinite-dimensional row vectors for 3D-vector-valued basis functions. We note here that all the fields expressed in the spherical basis can be easily converted to another coordinate system, such as the cartesian reference frame $(O,x,y,z)$ in Fig.~\ref{fig1}, simply with a suitable conversion matrix (the conversion to cartesian coordinates is given in Appendix~\eqref{app:A} for reference); this transformation will be left implicit in some formulas below to avoid cluttering already-lengthy formulas.

Given the linearity of the Helmholtz equation governing the scattering problem, the expansion coefficients of the incident field, $\widetilde{\mb{a}}$, and of the scattered field, ${\mb{f}}$, obey a linear relationship:
\begin{equation}
\mb{f}=\mb{T}\widetilde{\mb{a}}, \label{eq:T}
\end{equation} 
where $\mb{T}$ is the so-called \enquote{transition} or \enquote{transfer} matrix (\tmatrix\ for short). This quantity is an inherent property of the scatterer at a given frequency, and is independent of the incident field. It is this characteristic which leads to analytical formulas for orientation-averaged properties~\cite{StoutTransferMatrixMultipleScatterer2008}.

The elements of the \tmatrix\ can be determined through a variety of methods~\cite{Loke:2009aa,Fruhnert:2017aa,Reid:aa}; for simple shapes the Extended Boundary Condition Method, obtained by enforcing the boundary conditions on the scatterer's surface, is arguably the most accurate and efficient~\cite{somerville2016smarties}. In particular, for a homogeneous, spherical scatterer-centred at the origin, $\mb{T}$ is a diagonal matrix whose elements coincide with the Mie theory. For a system composed of several scatterers, the collective \tmatrix\ can be built up from the individual \enquote{one-body \emph{T}-matrices} of a single scatterer~\cite{StoutTransferMatrixMultipleScatterer2002, MishchenkoTravisLacisMain}. Following the notations of Ref.~\cite{StoutTransferMatrixMultipleScatterer2002} and Ref.~\cite{MishchenkoTravisLacisMain} we briefly recapitulate this \enquote{multi-scatterer \tmatrix} method.
\subsection{\texorpdfstring{The \tmatrix\ method for \emph{N} scatterers}{The T-matrix method for N scatterers}}
We now consider the situation depicted in Fig.~\ref{fig1}, where $N$ particles are present. With $\mathbf{r}$ we denote the position vector in the global \enquote{cluster} reference frame. The position of the centre of the $j$\textsuperscript{th} particle, $\mathbf{r}_0^{(j)}$, induces the position vector in this particle's frame, $\mathbf{r}_j = \mathbf{r} - \mathbf{r}_0^{(j)}$. 

The one-body \tmatrix\ of the $j$\textsuperscript{th} particle is denoted $T_1^{(j)}$. To keep the formulas relatively concise we assume in the following that each scatterer's \tmatrix\ is provided in the orientation of the common reference frame; in practice, this means that the \tmatrix\ of nonspherical particles, typically calculated in a high-symmetry orientation ($z$ axis coinciding with the symmetry axis for axisymmetric particles such as spheroids), needs to be rotated via the standard Wigner D-matrices~\cite{MishchenkoTravisLacisMain}.

In a system composed of $N$ scatterers, the net field exciting each particle is composed of the original incident field, plus the field scattered by other particles. If we denote by $e_N^{(j)}$ the coefficients of the electric field exciting particle $j$, and $f_N^{(j)}$ the scattered field coefficients from particle $j$,
\begin{align}
&f_N^{(j)} = T_1^{(j)} e_N^{(j)}, \label{eq:fN}\\
&e_N^{(j)} = J^{(j,0)} \Wa + \sum_{\mathclap{l=1, l\neq j}}^{N} H^{(j,l)} f_N^{(l)}. \label{eq:en}
\end{align}
where the regular translation matrix $J^{(j,0)}$ transforms the vector of incident field coefficients from the global origin ($\mb{r}_0$) to the position $\mb{r}_j$. The irregular translation matrices $H^{(j,l)}$ convert from an irregular VSWF expansion at $\mb{r}_l$ to a regular VSWF expansion at $\mb{r}_j$. \\
We can also formally consider the expansion coefficients $f_N^{(j)}$ of the field scattered by the $j$\textsuperscript{th} particle in response to the incident field $\Wa$, now in the presence of all other particles. Specifically, we define the $N$-body \tmatrix\ $T_N^{(j)}$ of the $j$\textsuperscript{th} particle to relate $f_N^{(j)}$ with $\Wa$ through  
\begin{align}
	f_N^{(j)} &= \qquad\quad T_N^{(j)} J^{(j,0)} \Wa,\label{eq:fN2}\\
		&\stackrel{\mathclap{\text{inc. plane waves}}}{=} \qquad\quad T_N^{(j)} \brr{e^{i\mathbf{k}\cdot \mathbf{r}_0^{(j)}} \Wa}.\label{eq:Tincplane}
\end{align}

These $N$-body \emph{T}-matrices can be obtained by replacing Eqs.~\eqref{eq:fN2} and \eqref{eq:en} in Eq.~\eqref{eq:fN} and solving the following coupled system of equations
\begin{equation}
	T_N^{(j)} = T_1^{(j)} \brr{I + \sum_{\mathclap{l=1,l\neq j}}^{N} H^{(j,l)} T_N^{(l)} J^{(l,j)}}.\label{eq:TN}
\end{equation}
where $I$ is the identity matrix. 
Equation~\eqref{eq:TN} can be used to implicitly define scatterer centred $N$-body \emph{T}-matrices $T_N^{(j,l)}$ \cite{Mackowski1994SphereClusters,StoutTransferMatrixMultipleScatterer2002} by setting 
\begin{equation}
T_N^{(j)}=\sum_{l=1}^N T_N^{(j,l)}J^{(l,j)}.\label{eq:TN2}
\end{equation}
In this way, $T_N^{(j,l)}$ described the full $N$-body contribution of particle $l$ to the \tmatrix\ of particle $j$.

After calculation of the one-body \emph{T}-matrices, by replacing Eqs.~\eqref{eq:TN2} and \eqref{eq:fN2} in Eq.~\eqref{eq:Esca} for a system of $N$ particles, the scattered electric field can be expressed as
\begin{equation}
\mb{E}_{\text{sca}}(k\mb{r})=\sum_{j=1}^N\sum_{l=1}^N \bPsi(k\mb{r}) T_N^{(j,l)} J^{(l,0)}\Wa.
\end{equation}
A great advantage of this particle-centred formalism is that the fields may be evaluated even inside the particle cluster, which is not possible when the whole cluster is described by a single global \tmatrix~\cite{StoutTransferMatrixMultipleScatterer2008}. 

We can now proceed to evaluate $\ldoc\propto\Im(\mb{E}^*\cdot\mb{B})$ from the knowledge of the local fields. First, according to the Maxwell equation $\mb{B} = -i\omega^{-1} \nabla \times \mb{E}$, and using Eq.~\eqref{eq:E-field} and the relation between $\widetilde{\mb{M}}$ and $\widetilde{\mb{N}}$~\cite{MackowskiMishchenko96}, we have $\mb{B}_\text{inc}=-ik\omega^{-1}\sum[\Wa_{1,nm}  \widetilde{\mb{N}}_{nm}+\Wa_{2,nm} \widetilde{\mb{M}}_{nm}]$. We re-write $\mb{B}$ in the compact form ($\mb{B}_{\text{inc}}(k\mb{r})=\WbPhi(k\mb{r})\widetilde{\mb{a}}$) analogous to Eq.~\eqref{eq:E-field2} for the electric field by defining $\WbPhi= [\widetilde{\mb{N}} \quad \widetilde{\mb{M}} ] $ for the incident field and similarly ${\mb{\Phi}}= [{\mb{N}} \quad {\mb{M}} ] $ for the scattered magnetic field. Expanding Eq.~\eqref{eq:OC1} into incident and scattered fields as $\ldoc  = \frac{-\omega \varepsilon_{0}}{2}\Im \left(\tb{E}_\text{inc}^*\tb{B}_\text{inc}+ \tb{E}_\text{inc}^*\tb{B}_\text{sca} + \tb{E}_\text{sca}^*\tb{B}_\text{inc}+\tb{E}_\text{sca}^*\tb{B}_\text{sca}\right)$ and inserting the corresponding expressions yields
\begin{equation}
\begin{split}
\ldoc= & \frac{k \varepsilon_{0}E_0^2}{2}\Re\Bigg[\Wa^\dagger \WbPsi^\dagger(k\mb{r}){\WbPhi}(k\mb{r}) \Wa +\\
{}& \sum_{j=1}^{N}\sum_{l=1}^{N}\Wa^\dagger {\WbPsi}^\dagger (k\mb{r})\bPhi(k\mb{r}_j){T}_N^{(j,l)}{J}^{(l,0)}\Wa + \\
{}& \sum_{j=1}^{N}\sum_{l=1}^{N}\Wa^\dagger J^{(0,l)}{T^\dagger}_N^{(j,l)}\bPsi^\dagger(k\mb{r}_j) \WbPhi(k\mb{r}) \Wa +\\
{}& \sum_{j=1}^{N}\sum_{l=1}^{N}\sum_{i=1}^{N}\sum_{k=1}^{N}\Wa^\dagger J^{(k,l)}{T^\dagger}_N^{(j,l)}\bPsi^\dagger(k\mb{r}_j) \bPhi(k\mb{r}_i) {T}_N^{(i,k)}\Wa\Bigg] \label{eq:OC}
\end{split}
\end{equation}
where $E_0$ is the incident electric field's amplitude and $\Re$ stands for \enquote{real part}. Note that each of the four terms is of the form $\Wa^\dagger \mb{M} \Wa$; this will be important in the derivation of the orientation-averaged $\langle\ldoc\rangle$ below.

We can simplify Eq.~\eqref{eq:OC} by translating the incident field to the centre of each particle via $J^{(l,0)}$ in the second and third terms of Eq.~\eqref{eq:OC}, and noting that for plane wave illumination $J^{(l,0)}=e^{i\mathbf{k}\cdot \mathbf{r}_0^{(l)}}I$, such that
\begin{equation}
\WbPsi(k\mb{r})\equiv\WbPsi(k\mb{r}_l)J^{(l,0)}, \quad 
\WbPhi(k\mb{r})\equiv\WbPhi(k\mb{r}_l)J^{(l,0)}.
\end{equation}    
Introducing these relations into Eq.~\eqref{eq:OC} leads to
\begin{equation}
\begin{split}
	\ldoc= & \frac{k \varepsilon_{0}E_0^2}{2}\Re\Bigg[\Wa^\dagger \WbPsi^\dagger(k\mb{r})\WbPhi(k\mb{r}) \Wa+ \\
	{}& \sum_{j=1}^{N}\sum_{l=1}^{N}\Wa^\dagger \WbPsi^\dagger (k\mb{r}_l)\bPhi(k\mb{r}_j){T}_N^{(j,l)}\Wa +\\
	{}& \sum_{j=1}^{N}\sum_{l=1}^{N}\Wa^\dagger{T^\dagger}_N^{(j,l)}\bPsi^\dagger(k\mb{r}_j) \widetilde {\bPhi}(k\mb{r}_l) \Wa +\\                                                             {}& \sum_{j=1}^{N}\sum_{l=1}^{N}\sum_{i=1}^{N}\sum_{k=1}^{N}\Wa^\dagger J^{(k,l)}{T^\dagger}_N^{(j,l)}\bPsi^\dagger(k\mb{r}_j) \bPhi(k\mb{r}_i) {T}_N^{(i,k)}\Wa\Bigg] \label{eq:OC2}
\end{split}
\end{equation}
where the addition rule of the translation matrices $I=J^{(l,0)}J^{(0,l)}$ is used. 

This expression for $\ldoc$ makes no assumption on the polarisation state of the incident field. We now turn to the description of optical chirality under illumination with circularly-polarised plane waves.

\subsection{Helicity versus parity bases for VSWFs}
In the standard basis of transverse-electric and transverse-magnetic VSWFs, the \tmatrix\ of Eq.~\eqref{eq:T} is usually written in the $2\times 2$ block-matrix form
\begin{align}
	\begin{bmatrix}
		T_{11} & T_{12}\\
		T_{21} & T_{22}
	\end{bmatrix}
	\begin{bmatrix}
		\Wa_1\\
		\Wa_2
	\end{bmatrix}=
	\begin{bmatrix}
		f_1\\
		f_2
	\end{bmatrix}
	\label{eq:T-parity}
\end{align}
where subscripts 1 and 2 refer to electric and magnetic multipolar contributions, respectively. The sub-matrices $T_{11}$ and $T_{22}$ describe the coupling of electric-electric and magnetic-magnetic multipolar components, while $T_{12}$ and $T_{21}$ describe cross-coupling of electric and magnetic multipolar components.

For circularly-polarised light, it is more useful to consider a different basis of VSWFs~\cite{OrientationAveragedMetaMaterials}, 
\begin{align}
	\begin{bmatrix}
		T_{\scriptscriptstyle LL} & T_{\scriptscriptstyle LR} \\
		T_{\scriptscriptstyle RL} & T_{\scriptscriptstyle RR}
	\end{bmatrix}
	\begin{bmatrix}
		\Wa_{\scriptscriptstyle L} \\
		\Wa_{\scriptscriptstyle R}
	\end{bmatrix}=
	\begin{bmatrix}
		f_{\scriptscriptstyle L} \\
		f_{\scriptscriptstyle R}
	\end{bmatrix}
	\label{eq:T-helicity}
\end{align} 
where the subscripts ($R$) and ($L$) refer to right and left circularly polarised light. This matrix describes the scattering of circularly-polarised incident fields in the helicity basis. The transformation from the parity basis can be obtained via the helicity operator ($\Lambda= \frac{\nabla \times}{k}$) and leads to,
\begin{subequations}
	\begin{align}
		& \mb{\helB}_{R,nm}=\frac{1}{\sqrt{2}}(\mb{M}_{nm}-\mb{N}_{nm}),\quad \Lambda\mb{\helB}_{R,nm} =-\mb{\helB}_{R,nm} \\
		& \mb{\helB}_{L,nm}=\frac{1}{\sqrt{2}}(\mb{M}_{nm}+\mb{N}_{nm}), \quad \Lambda\mb{\helB}_{L,nm}=\mb{\helB}_{L,nm}
	\end{align}.
	\label{eq:helical bases}
\end{subequations}  
%
%where subscripts $(R,L)$ refer to the helicity of the vector $%\mb{\helB}$ in this new basis.

These definitions lead to the following relation between the \tmatrix\ components in parity and helicity bases
\begin{align}
	\begin{bmatrix}
		T_{\scriptscriptstyle LL} & T_{\scriptscriptstyle LR} \\
		T_{\scriptscriptstyle RL} & T_{\scriptscriptstyle RR}
	\end{bmatrix}
	=\frac{1}{2}\begin{bmatrix}
		I & I \\
		I & -I
	\end{bmatrix}
	\begin{bmatrix}
		T_{11} & T_{12}\\
		T_{21} & T_{22}
	\end{bmatrix}
	\begin{bmatrix}
		I & I \\
		I & -I
	\end{bmatrix}, \label{eq:Tcp2}
\end{align}
where $I$ is the identity matrix with the same size as the 4 matrix blocks ($T_{11}$, etc.). 
Consequently, electric and magnetic fields in the helicity form become
\begin{align}
	&\mb{E}_{\text{tot}}=E_0 \left(\widetilde{\boldsymbol{\Gamma}}(k\mb{r})\Wa +\sum_{j=1}^N \boldsymbol{\Gamma}(k\mb{r}_j)T_N^{(j)} J^{(j,0)}\Wa\right),  \label{eq:Etotal}\\
	&\mb{B}_{\text{tot}}=E_0\frac{-ik}{\omega} \left(\widetilde{\boldsymbol{\Delta}}(k\mb{r})\Wa +\sum_{j=1}^N \boldsymbol{\Delta}(k\mb{r}_j)T_N^{(j)} J^{(j,0)}\Wa\right),
	\label{eq:Btotal}
\end{align}
where we introduce $\widetilde{\boldsymbol{\Gamma}}=[\widetilde{\mb{\helB}}_{L,nm} \quad \widetilde{\mb{\helB}}_{R,nm}]$, $\widetilde{\boldsymbol{\Delta}}=[\widetilde{\mb{\helB}}_{L,mn} \quad -\widetilde{\mb{\helB}}_{R,mn}]$, and the same convention for $\boldsymbol{\Gamma}$ and $\boldsymbol{\Delta}$ (composed of irregular VSWFs). Note that the $T_N^{(j)}$ as well as $J^{(j,0)}$ in Eq.~\eqref{eq:Etotal} need to be also in the helicity basis, following the same transformation as Eq.~\eqref{eq:Tcp2}. 

With the general expression for $\ldoc$ and the transformations between parity and helicity bases, we can now proceed to calculate the orientation-averaged $\ldoc$ for circularly polarised incident light.
\subsection{\texorpdfstring{Orientation-averaging of $\ldoc$}{Orientation-averaging of LDOC}}
A distinctive advantage of the \tmatrix\ framework is that any variation of the incident field is captured in the $\Wa$ expansion coefficients, while the \tmatrix\ itself is unchanged. From this property and the orthonormality of VSWFs over the 2-sphere, the orientation average of any optical property that can be written as a bilinear product of the form $ \Wa^\dagger \mb{M} \Wa$ reduces to a simple trace formula~\cite{StoutTransferMatrixMultipleScatterer2008,Suryadharma18},
\begin{equation}
\langle \Wa^\dagger \mb{M} \Wa\rangle=4\pi \Tr(\mb{M}) \label{eq:Tr}
\end{equation}  
where $\Tr$ stands for \enquote{trace of} and $\mb{M}$ is any bilinear operator whose matrix coefficients are independent of the incident field. This formula can be directly applied to evaluate the orientation-average of Eq.~\eqref{eq:OC2}. We consider an incident field described in the helicity basis, and restrict the coefficients to a single helicity (corresponding to either left or right polarisation). This corresponds to $\Wa=[0 \quad a_{\scriptscriptstyle R}]^t$ for $R$ polarisation, and $\Wa=[a_{\scriptscriptstyle L} \quad 0]^t$ for $L$ polarisation. The result in both cases contains four different terms (See Appendix~\eqref{app:B} for more details),

\begin{equation}
\langle \ldoc \rangle=2\pi k \varepsilon_{0}E_0^2\,\Re \left(A_0+B_0+C_0+D_0\right)
\label{eq:avg}
\end{equation}
with,
\begin{widetext}
For $R$ polarisation:
\begin{equation}
\begin{split}
	A^\text{(R)}_0= & {-1}/{4\pi}                                                                                                                                                                                                        \\
	B^\text{(R)}_0= & \Tr\left( \sum_{j=1}^{N}\sum_{l=1}^{N}\widetilde{\mb{Z}}_{\scriptscriptstyle R}^{\dagger}(k\mb{r}_l)\left[ \mb{Z}_{\scriptscriptstyle L}(k\mb{r}_j)T_{\scriptscriptstyle LR}^{(j,l)}-\mb{Z}_{\scriptscriptstyle R}(k\mb{r}_j)T_{\scriptscriptstyle RR}^{(j,l)}\right] \right)                                  \\
	C^\text{(R)}_0= & \Tr\left( \sum_{j=1}^{N}\sum_{l=1}^{N}\left[-T_{\scriptscriptstyle LR}^{\dagger(j,l)}\mb{Z}_{\scriptscriptstyle L}^{\dagger}(k\mb{r}_j)-T_{\scriptscriptstyle RR}^{\dagger(j,l)}\mb{Z}_{\scriptscriptstyle R}^{\dagger}(k\mb{r}_j)\right] \widetilde{\mb{Z}}_{\scriptscriptstyle R}(k\mb{r}_l)\right)             \\
	D^\text{(R)}_0= & \Tr \left( \sum_{j=1}^{N}\sum_{l=1}^{N}\sum_{i=1}^{N}\sum_{k=1}^{N}J_{\scriptscriptstyle RR}^{(k,l)}\left(T^{\dagger(j,l)}_{\scriptscriptstyle LR}\mb{Z}_{\scriptscriptstyle L}^{\dagger}(k\mb{r}_j)+T^{\dagger(j,l)}_{\scriptscriptstyle RR}\mb{Z}_{\scriptscriptstyle R}^{\dagger}(k\mb{r}_j)\right)\left(\mb{Z}_{\scriptscriptstyle L}(k\mb{r}_i)T_{\scriptscriptstyle LR}^{(i,k)}-\mb{Z}_{\scriptscriptstyle R}(k\mb{r}_i)T_{\scriptscriptstyle RR}^{(i,k)}\right)\right).
\end{split} \label{eq:finalR}
\end{equation}
Notice that the sum of $B^\text{(R)}_0$ and $C^\text{(R)}_0$ simplifies to,
\begin{equation}
\Re(B^\text{(R)}_0+C^\text{(R)}_0)=-2 \Tr\left(\sum_{j=1}^{N}\sum_{l=1}^{N}\widetilde{\mb{Z}}_{\scriptscriptstyle R}^{\dagger}(k\mb{r}_l)\mb{Z}_{\scriptscriptstyle R}(k\mb{r}_j)T_{\scriptscriptstyle RR}^{(j,l)}\right).
\end{equation}
The corresponding formulas for $L$ polarisation read,
\begin{equation}
\begin{split}
	A^\text{(L)}_0= & {+1}/{4\pi}                                                                                                                                                                                                        \\
	B^\text{(L)}_0= & \Tr\left( \sum_{j=1}^{N}\sum_{l=1}^{N}\widetilde{\mb{Z}}_{\scriptscriptstyle L}^{\dagger}(k\mb{r}_l)\left[ \mb{Z}_{\scriptscriptstyle L}(k\mb{r}_j)T_{\scriptscriptstyle LL}^{(j,l)}-\mb{Z}_{\scriptscriptstyle R}(k\mb{r}_j)T_{\scriptscriptstyle RL}^{(j,l)}\right] \right)                                  \\
	C^\text{(L)}_0= & \Tr\left( \sum_{j=1}^{N}\sum_{l=1}^{N}\left[T_{\scriptscriptstyle LL}^{\dagger(j,l)}\mb{Z}_{\scriptscriptstyle L}^{\dagger}(k\mb{r}_j)+ T_{\scriptscriptstyle RL}^{\dagger(j,l)}\mb{Z}_{\scriptscriptstyle R}^{\dagger}(k\mb{r}_j)\right] \widetilde{\mb{Z}}_{\scriptscriptstyle L}(k\mb{r}_l)\right)             \\
	D^\text{(L)}_0= & \Tr \left( \sum_{j=1}^{N}\sum_{l=1}^{N}\sum_{i=1}^{N}\sum_{k=1}^{N}J_{\scriptscriptstyle LL}^{(k,l)}\left(T^{\dagger(j,l)}_{\scriptscriptstyle LL}\mb{Z}_{\scriptscriptstyle L}^{\dagger}(k\mb{r}_j)+T^{\dagger(j,l)}_{\scriptscriptstyle RL}\mb{Z}_{\scriptscriptstyle R}^{\dagger}(k\mb{r}_j)\right)\left(\mb{Z}_{\scriptscriptstyle L}(k\mb{r}_i)T_{\scriptscriptstyle LL}^{(i,k)}-\mb{Z}_{\scriptscriptstyle R}(k\mb{r}_i)T_{\scriptscriptstyle RL}^{(i,k)}\right)\right),
\end{split} 
\end{equation}
where the sum of $B^\text{(L)}_0$ and $C^\text{(L)}_0$ simplifies to,
\begin{equation}
\Re(B^\text{(L)}_0+C^\text{(L)}_0)= 2 \Tr\left(\sum_{j=1}^{N}\sum_{l=1}^{N}\widetilde{\mb{Z}}_{\scriptscriptstyle L}^{\dagger}(k\mb{r}_l)\mb{Z}_{\scriptscriptstyle L}(k\mb{r}_j)T_{\scriptscriptstyle LL}^{(j,l)}\right). \label{eq:finalL}
\end{equation}
\end{widetext} 
In some circumstances we may also be interested in averaging over both possible states of circular polarisation, with $\tfrac{\langle \ldoc \rangle^\text{(R)} + \langle \ldoc \rangle^\text{(L)}}{2}$, which should vanish everywhere for achiral cluster geometries. 

The terms $A_0^{(R)}$ and  $A_0^{(L)}$ of Eqs.~\eqref{eq:finalR} and \eqref{eq:finalL}, respectively, are equal and opposite and correspond to the results for circularly-polarised plane waves. This highlights the fact that in the absence of a scatterer, and for linearly polarised light, $\ldocoa$ vanishes identically: $a_{\text{R}}$ and $a_{\text{L}}$ are of the same magnitude, only  $A_0^{(R)}$ and  $A_0^{(L)}$ contribute to $\ldocoa$, and cancel each other. This is also clear from Eq.~\eqref{eq:OC2} where the first term contains $\WbPsi^\dagger(k\mb{r})\WbPhi(k\mb{r})$. An integration over the 2-sphere for linearly-polarised incoming plane waves will necessarily make this term vanish by orthogonality of different VSWFs.

The value of $\ldoc$ is often normalised with respect to the value of circularly-polarised plane waves with unit incident field, $\ldoc=\pm k\varepsilon_0/2$, and we introduce
\begin{equation}
\ldocoabar = \frac{2}{k\varepsilon_0E_0^2} \ldocoa
\end{equation}
for easier comparisons in the examples below.

% Key formulas
% \newpage
%
\section{Application to model nanostructures}
We have applied the newly-derived formula Eqs.~\eqref{eq:finalR}--\eqref{eq:finalL} to a range of cluster geometries, with both spherical and nonspherical scatterers, metallic and dielectric. Our custom-built computer program for the scatterer-centred superposition \tmatrix\ method is based on the algorithms described in Refs.~\cite{MackowskiMishchenko96,StoutTransferMatrixMultipleScatterer2002} and has been tested against FEM simulations~\cite{Schebarchov:2019aa}. The validity of Eqs.~\eqref{eq:finalR} and \eqref{eq:finalL} was confirmed by comparing the results to orientation-averaged values of $\ldocoa$ computed by numerical integration of Eq.~\eqref{eq:eulerOA} over the full solid angle. The \tmatrix\ of spheres is obtained from Mie theory, while for spheroids we use the \textsc{smarties} program~\cite{somerville2016smarties}. Numerical implementations of the \tmatrix\ method require us to truncate the series of VSWFs to a maximum multipolar order $N_{\text{max}}$; in our calculations a value of $N_{\text{max}} = 8$ was used, which was sufficient to ensure convergence of the results. 

The following examples were chosen to illustrate the results in simple geometries but the method is equally applicable to more complex structures, for which an in-depth study is beyond the scope of this manuscript.

% illustrations on relevant examples

%
\subsection{Monomer and dimer of silver and silicon nanospheres}
%

% - how does the lodoc vary in space?
% - hot-spots near a dimer?
% - magnetic mode, metal vs dielectric particles?
% - surface averaging vs pointwise?
% - mention kerker condition

% {\it
% illustrations on relevant examples, to answer the following:
% 
% \begin{itemize}
% \item
%   how does the lodoc vary in space?
% \item
%   hot-spots near a dimer?
% \item
%   magnetic mode, metal vs dielectric particles?
% \item
%   surface averaging vs pointwise?
% \item
%   mention kerker condition
% \end{itemize}
% }

\begin{figure}
	\centering
\includegraphics[width=\columnwidth]{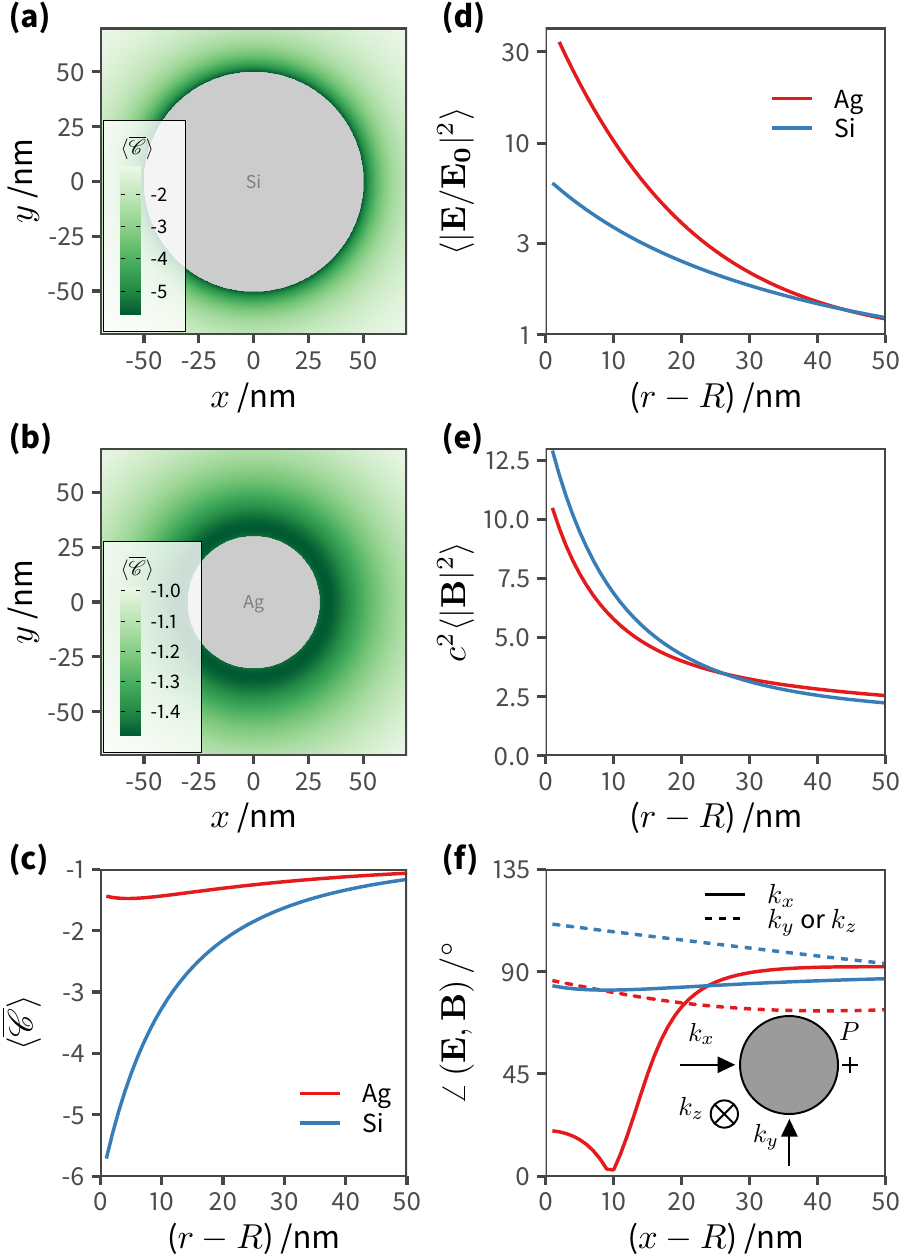}
	\caption{(a, b) 2D colour maps of the normalised orientation-averaged local degree of optical chirality $\ldocoabar$ around a single Si sphere with radius $R=\SI{50}\nm$ excited at $\lambda =\SI{480}\nm$ (a), and a Ag sphere with $R=\SI{30}\nm$ at $\lambda =\SI{468}\nm$ (b). Note that the patterns are spherically symmetric. (c, d, e) Corresponding plot of the radial dependence of $\ldocoabar$, $\langle |\mb{E}|^2 \rangle$, $\langle |\mb{B}|^2 \rangle$ with origin at the surface of the sphere. Note that $\langle |\mb{E}|^2 \rangle$ is displayed on a log scale to better compare the relative decay profiles, and that we scale the magnetic field by the speed of light $c=\SI{299792458}m/s$ to retain the same order of magnitude as the electric field. (f) Radial dependence of the phase angle between $\mb{E}$ and $\mb{B}$ at the location $P$ along the $x$ direction as shown in the inset, for three particular incidence directions, $k_x$ (solid line) and $k_y$ or $k_z$ (dashed line, both directions are equivalent).}
	\label{fig2}
\end{figure}

As the simplest case study, we start with a single nanosphere immersed in water (refractive index $n = 1.33$), and consider both a plasmonic (Ag) and a high-index dielectric (Si) spheres, with dielectric functions from Refs.~\cite{LeRuE08} and \cite{aspnes1983dielectric}, respectively. Silver nanoparticles have been used extensively in surface-enhanced spectroscopies such as SERS, while silicon nanospheres have been shown to support strong electric and magnetic resonances in the visible spectrum~\cite{evlyukhin2014optical,Zhang:2017aa,Ho:2017aa}. The radii are chosen as $30$\,nm for Ag and $50$\,nm for Si, to display a relatively high value of optical chirality. 

As expected the local degree of optical chirality reaches its highest value at the particle surface where electric and magnetic fields are strongest, and decays to the value of $\pm 1$ far from the scatterer, corresponding to the degree of optical chirality of a circularly polarised plane wave (Fig.~\ref{fig2}(a--b)). The distributions are identical for either $L$ or $R$ polarisation as the structure is achiral, with opposite sign; for simplicity we only present the results for $R$ polarisation. $\ldocoabar$ for Si is nearly 4 times higher at the surface than $\ldocoabar$ for the Ag sphere.

\begin{figure*}[!htpb]
	\centering
\includegraphics[width=\textwidth]{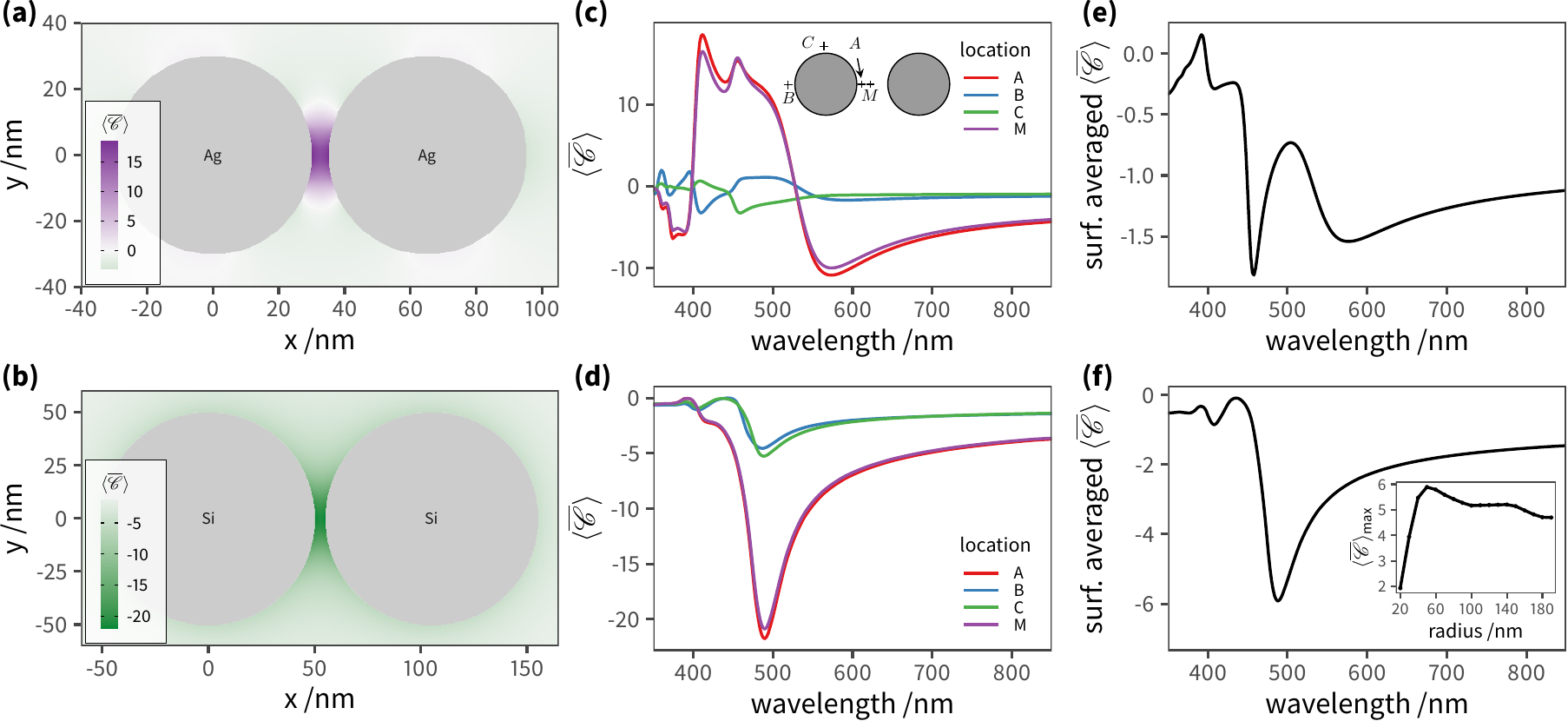}
	\caption{(a) Map of the normalised local degree of optical chirality $\ldocoabar$ for a dimer of silver spheres with radius $R=\SI{30}\nm$ and gap $\SI{5}\nm$, calculated on a 2D plane passing through the sphere centres. The wavelength of excitation is $\lambda =\SI{413}\nm$, corresponding to the peak extinction. (b) Same configuration as in (a) for a dimer of $\SI{50}\nm$-radius Si spheres with $\SI{5}\nm$ gap, illuminated at $\lambda =\SI{490}\nm$ (peak extinction). (c, d) Calculated $\ldocoabar$ spectra at four different locations around each dimer, as depicted in the inset. The point M is in the middle of the gap and points A, B, C are $\SI{0.1}\nm$ away of the nanoparticle's surface. (e, f) Surface-averaged of $\ldocoabar$ across the particles' surface for each dimer. The inset in (f) traces the maximum surface-averaged $\ldocoabar$ for a Si dimer with varying sphere radius ($\SI{20}\nm-\SI{200}\nm$) and constant gap ($\SI{5}\nm$).}
	\label{fig3}
\end{figure*}

Less intuitive is the decay of $\ldocoabar$ with distance from the particle surface, which is very different for both cases (Fig.~\ref{fig2}(c)), and does not directly correlate with the decay profile of the electric field intensity $|\mb{E}|^2$, more familiar in this context. The degree of optical chirality depends on both electric and magnetic fields, and therefore on the relative phase of each of their vector components. A general case could be difficult to interpret, but in this highly-symmetric configuration we can gain some insight by looking at the relative phase of $\mb{E}$ and $\mb{B}$, calculated as $\angle\left(\mb{E},\mb{B}\right)=\cos^{-1}\tfrac{\Re(\mb{E}^*.\mb{B})}{|\mb{E}||\mb{B}|}$, at a fixed location and for a given direction of incidence. The relative phases versus distance from the particle surface are plotted in Fig.~\ref{fig2}(f) for three perpendicular incident directions, with the observation point $P$ along the $x$ axis. We observe that although the electric field is more intense near the Ag sphere (with nearly equal magnetic fields (Fig.~\ref{fig2}(e))), the phase difference between $\mb{E}^*$ and $\mb{B}$ for a silver sphere is more variable with the incident direction. Moreover, the phase difference for a silicon sphere is nearly 90 degrees, which in turn leads to the comparatively higher value of $\ldocoabar$. These considerations confirm the potential benefit of using high-index dielectric resonators instead of plasmonic particles~\cite{Ho:2017aa,Zhang:2017aa,Yao:2018aa}, as they can offer better control over magnetic fields~\cite{sym11091113} and preserve the helicity of the incident light~\cite{Solomon:2019aa,Lasa-Alonso:2020aa}

The next obvious structure to consider is a dimer of nanospheres, the simplest multi-particle configuration which nevertheless has served as the workhorse of many SERS studies, as the gap between nanospheres can support highly-localised electromagnetic \enquote{hotspots}. In Fig.~\ref{fig3} we consider the orientation-averaged $\ldocoabar$ for both Si and Ag dimers, with the dimer axis along $x$ and a $\SI{5}{\nano\metre}$ gap. At the hotspot for the Si dimer the absolute value of $\ldocoabar$ reaches $21.8$, and $18.5$ for the Ag dimer, and for the latter the helicity has changed sign with respect to the incident right-handed circularly-polarised light.

In this configuration the value of $\ldocoabar$ varies with position on either sphere. Since analytes cannot generally be placed at will in the optimum region~\cite{Le-Ru:2011aa}, we present the spatial variation at three particular locations on the sphere (by symmetry both spheres in a dimer present the same pattern)(Fig.~\ref{fig3}(c,d)). We also present in Fig.~\ref{fig3}(e,f) the surface-averaged value of $\ldocoabar$, evaluated numerically with a 38-point Lebedev spherical quadrature rule. We observe that although the maximum absolute values of $\ldocoabar$ are comparable in both cases, the surface-average is considerably weaker for the Ag dimer than for the Si dimer ($-1.7$ and $-6$, respectively), due to the sign change of $\ldocoabar$ across the Ag dimer. 

Since the formulas \eqref{eq:finalR}--\eqref{eq:finalL} are analytical and can be evaluated with negligible computational overhead once the particle-centred \emph{T-}matrices have been obtained, we can explore the effect of various parameters on the value and distribution of $\ldocoa$, as a first step toward identifying optimised structures for practical applications. Such an example is shown in inset of Fig.~\ref{fig3}(f), for the maximum value of the surface-averaged $\ldocoabar$ as a function of particle radius, for a fixed $\SI{5}{\nano\metre}$ gap. By increasing the radius of the Si nanospheres the magnitude of both electric and magnetic fields increases, and up to $\SI{50}\nm$ the phase difference between them remains nearly $\pi/2$; beyond $\SI{50}\nm$ however the phase difference varies more and results in a decrease of the surface-averaged $\ldocoabar$.      

%
% \subsection{Chiral vs achiral nanostructures}
%

% questions to address: 
% - does the chirality of the nanostructure affect the orientation-averaged lodoc?
% - if so, how?
% - orientation-averaging vs full averaging, including polar

% {\it
% questions to address:
% 
% \begin{itemize}
% \item
%   does the chirality of the nanostructure affect the
%   orientation-averaged lodoc?
% \item
%   if so, how?
% \item
%   orientation-averaging vs full averaging, including polar
% \end{itemize}
% }

%
\subsection{Analytical vs numerical orientation-averaging}
\begin{figure}[!htpb]
	\centering
\includegraphics[width=\columnwidth]{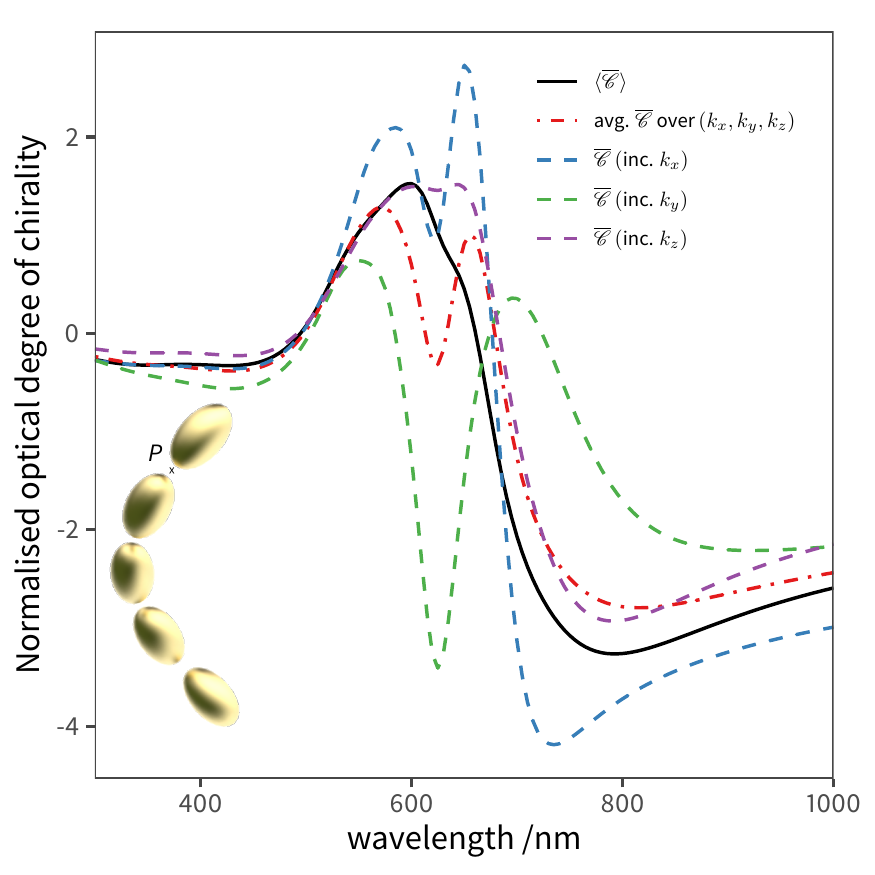}
	\caption {Predicted spectrum of the local degree of optical chirality at the point $P$ half-way between the last two particles, as depicted in the inset. The structure consists of a helix of five prolate gold spheroids, with helix axis $z$, radius $\SI{100}\nm$, pitch $\SI{700}\nm$, angular-step $\delta = \pi/4$. The spheroids are oriented along the helix, and have semi-axes $a=b=\SI{30}\nm$ and $c = \SI{50}\nm$. The structure is illuminated with a right circularly-polarised plane wave along the $x$ direction (blue dashed line), $y$ direction (green dashed line), or $z$ direction (purple dashed line). The average over these three incident directions (red dashed line) is compared to the rigorous orientation-averaged value $\ldocoabar$ (solid black line).}
	\label{fig4}
\end{figure}
Our last example is a helix of gold nanorods, depicted in Fig.~\ref{fig4}. Elongated nanoparticles have been shown to amplify the far-field circular dichroism in chain-like~\cite{Severoni:2020aa} or helical structures~\cite{Guerrero-Martainez:2011ab}, and changing the particles' aspect ratio provides a convenient means to tune the resonance position. This system also provides a useful test for our numerical implementation of the formalism, requiring \emph{T-}matrices of spheroidal particles~\cite{somerville2016smarties}, and also the rotation of each individual particle. 

The helix is composed of five identical Au spheroids with semi-axes $a = \SI{30}\nm , c = \SI{50}\nm$; the radius of the helix is $\SI{100}\nm$ and pitch $\SI{700}\nm$, and the angular step between each particle is $\delta = \pi/4$. The orientation-averaged value of $\ldocoabar$ at a particular \enquote{hotspot} position $P$ between the last two particles is compared in Fig.~\ref{fig4} to the value obtained for three particular directions of incidence ($k_x, k_y, k_z$, respectively). The value of $\ldocbar$ is very different for the three incident directions, and their \enquote{naive} averaging provides an unreliable estimate of the true orientation-averaged value $\ldocoabar$. This points to the importance of proper orientation-averaging when comparing simulations to experiments where samples are randomly-oriented with respect to the incident light. The size and symmetry of the cluster as well as the strength of the interparticle coupling will contribute to the difficulty in estimating numerically the orientation-average with discrete simulations at specific directions of incidence. The analytical results presented here can serve as a useful benchmark to evaluate the number of quadrature points needed for a numerical evaluation of Eq.~\eqref{eq:eulerOA}. A comparison of spherical quadrature rules for both far-field and near-field quantities is beyond the scope of this study but will be presented elsewhere.

% point to address: 
% - show that analytical averaging is super useful
% - give an example where 3 directions is clearly not enough
% - hint at the follow-up paper studying quadrature

%
\section{Conclusion}
%
% \begin{itemize}\it
% \item What we've found. 
% \item How useful is it. 
% \item What else can be done with this. 
% \item Open questions/challenges
% \end{itemize}

We have derived analytical formulas for the orientation-averaged degree of optical chirality $\ldocoa$ around clusters of nanoparticles illuminated with circularly-polarised light, using the rigorous superposition-\tmatrix\ framework. The formulas, although lengthy, are readily implemented in software, and provide an efficient method to calculate rigorously-averaged results, in contrast to numerical estimates obtained from several directions of incidence, as is commonly done with other numerical methods. The local degree of optical chirality informs us on the subtle interplay between complex electric and magnetic fields, and how their combination may enhance the chiroptical response of molecules in the vicinity of a cluster of particles. We illustrated the behaviour of $\ldocoa$ by mapping its distribution around single spheres and dimers of Ag and Si, with less intuitive results compared to the better-known electric field intensity. The importance of the analytical formula for $\ldocoa$ was illustrated for a helix of elongated particles, where we show that a simple averaging of the results from three orthogonal directions provides a poor estimate of the true value. 

These few examples illustrate the broad range of studies that become tractable with an efficient and accurate formulas for $\ldocoa$, notably the search for optimal structures that maximise the chiroptical response of adsorbed species, either at a specific location in space, or as averages over the surface of the particles.

The \tmatrix\ framework used to arrive at these results is ideally-suited for many studies requiring  orientation-averaging, and we hope this paper serves to demonstrate its particular usefulness in the  context of near-field chiroptical properties where it has not received much attention to date, with most of the literature relying on purely numerical solutions such as FDTD, or coupled-dipole approximations with limited range of validity. Although the superposition \tmatrix\ method provides a rigorous solution to the light scattering problem, it also suffers some limitations, in particular where the circumscribed spheres of neighbouring particles overlap~\cite{Schebarchov:2019aa}. We hope this study will also trigger further work in improving the method and overcoming its current limitations~\cite{PhysRevA.96.033822}.

\section*{Acknowledgements}
The authors would like to thank Dmitri Schebarchov and Eric Le Ru for helpful discussions, the Royal Society Te Ap\=arangi for support through a Rutherford Discovery Fellowship (B.A.), and the MacDiarmid Institute for additional funding (A.F.N., S.S.).

\bibliography{references}
%
% \newpage

\appendix
\section{Vector Spherical Harmonics}\label{app:A}
%
% \subsection*{Vector Spherical Harmonics}\label{sec:VSWFandCo} 
%
The vector spherical harmonics ($\widetilde{\mb{M}},\widetilde{\mb{N}},{\mb{M}},{\mb{N}}$) are defined as follows
\begin{equation}
\begin{matrix}
\widetilde{\mb{M}}_{nm}(k\mb{r},\theta,\varphi) \\
{\mb{M}}_{nm}(k\mb{r},\theta,\varphi)
\end{matrix}
=\gamma_{nm} \nabla\times
\begin{pmatrix}
\mb{r}\widetilde{\xi}_{nm}(k\mb{r},\theta,\varphi) \\
\mb{r}{\xi}_{nm}(k\mb{r},\theta,\varphi)
\end{pmatrix},\\
\end{equation}
and
\begin{equation}
\begin{matrix}
\widetilde{\mb{N}}_{nm}(k\mb{r},\theta,\varphi) \\
{\mb{N}}_{nm}(k\mb{r},\theta,\varphi)
\end{matrix} = \frac{1}{k} \nabla \times
\begin{pmatrix}
\widetilde{\mb{M}}_{nm}(k\mb{r},\theta,\varphi) \\
{\mb{M}}_{nm}(k\mb{r},\theta,\varphi)
\end{pmatrix},
\end{equation}
where 
\begin{align}
&\begin{pmatrix}
\widetilde{\xi}_{nm}(k\mb{r},\theta,\varphi) \\
{\xi}_{nm}(k\mb{r},\theta,\varphi)
\end{pmatrix}
=\quad
\begin{pmatrix}
j_n(k\mb{r})\\
h_n^{(1)}(k\mb{r})
\end{pmatrix} Y_{nm}(\theta,\varphi)\label{eq:sphH}\\
&\gamma_{nm} =\left[ \frac{(2n+1)(n-m)!}{4\pi n (n+1)(n+m)!} \right].
\end{align}
In Eq.~\eqref{eq:sphH}, $j_n(k\mb{r})$ and $h_n^{(1)}(k\mb{r})$ are regular and irregular spherical Bessel functions, respectively and the spherical harmonics are defined as
\begin{equation}
Y_{nm}(\theta,\varphi) = P_n^m(\cos(\theta))e^{im\varphi},
\end{equation}
$P_n^m(\cos(\theta))$ are the associated Legendre functions, defined using the Condon–Shortley phase
\begin{equation}
P_n^m(x)=\frac{(-1)^m}{2^n n!}(1-x^2)^{m/2} \frac{d^{n+m}}{dx^{n+m}} (x^2-1)^{n}, m\geq 0
\end{equation}
and
\begin{equation}
P_n^{(-m)}= (-1)^m \frac{(n-m)!}{(n+m)!}P_n^m.
\end{equation}
\subsection*{Transformation to cartesian coordinates}
Since we seek vector fields expressed in the cartesian basis of the cluster reference frame, for the calculation of different terms of $\ldoc$ (Eqs.~\eqref{eq:finalR}-\eqref{eq:finalL}) we must convert electric and magnetic field components from a spherical to a cartesian basis. The matrix $C_j$ to perform this transformation for particle $j$ reads:
\begin{equation}
	C_j = \begin{pmatrix}
		\sin\theta_j\cos\varphi_j &  \cos\theta_j \cos\varphi_j & -\sin\varphi_j \\
		\sin\theta_j\sin\varphi_j &  \cos\theta_j \sin\varphi_j &  \cos\varphi_j  \\
		\cos\theta_j             & -\sin\theta_j           & 0
	\end{pmatrix}.
\end{equation}
$C_0$ is the corresponding matrix for the global \enquote{cluster} coordinate system. Note that the $C_j$ matrices are needed even if all particle frames are parallel, to transform field components from spherical unit vectors to cartesian ones.

% \section{A Quick Summary of Relevant $\mathbf{T}$-Matrix Methods}
%
\section{Derivation of Eq.~\eqref{eq:finalR}}\label{app:B}
We provide here a step-by-step derivation to arrive at equations~\eqref{eq:finalR} and \eqref{eq:finalL}. The key ingredient is to carefully keep track of the block matrix structure of the involved quantities in the helicity basis for VSWFs. For the sake of brevity, we will explicitly perform this derivation for right circularly-polarised light, Eq.~\eqref{eq:finalR}. The modifications needed to derive Eq.~\eqref{eq:finalL} are minimal.

As discussed in the main text, right circularly-polarised light is described with an expansion coefficient $\Wa^\text{R}$ that has the block structure $\Wa^\text{R}_{\scriptscriptstyle LR}=[0 \quad a_{\scriptscriptstyle R}]^t$ in the helicity basis (more generally, we use the $LR$ subscript to refer to helicity basis in the following). Then, starting from Eq.~\eqref{eq:OC2}, we convert each term to its helical basis representation and insert the expansion coefficient for $\Wa$. The resulting equation is the sum of four terms $\ldoc=\frac{k \varepsilon_{0}E_0^2}{2}\,\Re \left(A+B+C+D\right)$, where we named the terms in a way indicative of their contributions to the final results, and we derive them sequentially below.

First, notice that the basis transformation changing $\Wa$ to $\Wa_{\scriptscriptstyle LR}$ is simply given by
\begin{equation}
	\Wa_{\scriptscriptstyle LR} = \frac{1}{\sqrt{2}}\begin{bmatrix}
		I &I \\ I &-I
	\end{bmatrix}\Wa,
\end{equation}
which holds independent of whether we look at regular or irregular VSWFs. This transformation is self-adjoint, real, and involutive. 

	As a result, it is straightforward to get the precursor of $A^\text{(R)}_0$ from Eq.~\eqref{eq:OC2}:
	\begin{align}
		A^\text{(R)} &=(\Wa^\text{R})^\dagger \WbPsi^\dagger(k\mb{r})\WbPhi(k\mb{r}) \Wa^\text{R}, \\
		&= (\Wa^\text{R}_{\scriptscriptstyle LR})^\dagger \widetilde{\boldsymbol{\Gamma}}^\dagger(k\mb{r})\widetilde{\boldsymbol{\Delta}}(k\mb{r})  \Wa^\text{R}_{\scriptscriptstyle LR}.
	\end{align}
	At this point, we insert the definition of $\widetilde{\boldsymbol{\Gamma}}$ and $\widetilde{\boldsymbol{\Delta}}$. When doing this, it is useful to remember that to transpose a block matrix, one first treats the blocks as entries of a matrix, transposes this matrix as usual, and then additionally transposes each of the component blocks. This gives:
	\begin{equation}
	A^\text{(R)} =  \Wa_{\scriptscriptstyle R}^\dagger \widetilde{\mb{\helB}}_{\scriptscriptstyle R}^{\dagger} \left( -\widetilde{\mb{\helB}}_{\scriptscriptstyle L}\right)\Wa_{\scriptscriptstyle R}.
	% 	A^\text{(R)} = \Re\Bigg[\sum_{mn}\sum_{m'n'}(\Wa_{\text{R},mn})^\dagger \widetilde{\mb{\helB}}_{R,mn}^\dagger(-\widetilde{\mb{\helB}}_{R,m'n'}) \Wa_{\text{R},m'n'}\Bigg].
	\end{equation}
Orientation-averaging of this expression relies on the orthogonality relations between VSWFs on the sphere and must lead to a constant. The main difficulty is to keep track of the normalisation factors, but here we can also realise that the average value of $\ldocbar$ for the incident field must equal the value of a single circularly-polarised plane wave, which is independent of its direction of propagation. This gives $A^\text{(R)}_0 = -\tfrac{1}{4\pi}$.

This combination of matrix algebra before performing the orientation average repeats in the other terms of Eq.~\eqref{eq:finalR}. In the following we therefore focus on the matrix structure. For $B^\text{(R)}$, we write
	\begin{multline}
		B^\text{(R)}  = \\ 
		{}\quad	\sum_{j,l=1}^{N}[0, \Wa_{\scriptscriptstyle R}^\dagger] \begin{bmatrix}
			\widetilde{\mb{\helB}}_{\scriptscriptstyle L}^{l,\dagger} \\ \widetilde{\mb{\helB}}_{\scriptscriptstyle R}^{l,\dagger}
		\end{bmatrix} [\widetilde{\mb{\helB}}_{\scriptscriptstyle L}^j, -\widetilde{\mb{\helB}}_{\scriptscriptstyle R}^j]
		\begin{bmatrix}
			(T^{(l,j)}_N)_{\scriptscriptstyle LL} & (T^{(l,j)}_N)_{\scriptscriptstyle LR}\\ (T^{(l,j)}_N)_{\scriptscriptstyle RL} & (T^{(l,j)}_N)_{\scriptscriptstyle RR}
		\end{bmatrix}
		\begin{bmatrix}
			0\\\Wa_{\scriptscriptstyle R}
		\end{bmatrix}\\
	{}\quad = \sum_{j,l=1}^{N} \Wa_{\scriptscriptstyle R}^\dagger \widetilde{\mb{\helB}}_{\scriptscriptstyle R}^{l,\dagger} \left( \widetilde{\mb{\helB}}_{\scriptscriptstyle L}^j (T^{(l,j)}_N)_{\scriptscriptstyle LR} - \widetilde{\mb{\helB}}_{\scriptscriptstyle R}^j (T^{(l,j)}_N)_{\scriptscriptstyle RR} \right)\Wa_{\scriptscriptstyle R},
	\end{multline}
which, upon integration, leads to $B^\text{(R)}_0$.
	
In the expressions for $C^\text{(R)}$ we need to account for the transposition of block matrices:
	\begin{multline}
		C^\text{(R)}  =  \\
{}\quad\sum_{j,l=1}^{N}[0, \Wa_{\scriptscriptstyle R}^\dagger]\begin{bmatrix}
			(T^{(l,j)}_N)_{\scriptscriptstyle LL}^\dagger & (T^{(l,j)}_N)_{\scriptscriptstyle RL}^\dagger\\ (T^{(l,j)}_N)_{\scriptscriptstyle LR}^\dagger & (T^{(l,j)}_N)_{\scriptscriptstyle RR}^\dagger
		\end{bmatrix} \begin{bmatrix}
			\widetilde{\mb{\helB}}_{\scriptscriptstyle L}^{l,\dagger} \\ \widetilde{\mb{\helB}}_{\scriptscriptstyle R}^{l,\dagger}
		\end{bmatrix} [\widetilde{\mb{\helB}}_{\scriptscriptstyle L}^j, -\widetilde{\mb{\helB}}_{\scriptscriptstyle R}^j]
		\begin{bmatrix}
			0\\\Wa_{\scriptscriptstyle R}
		\end{bmatrix}\\
{}\quad		= -\sum_{j,l=1}^{N} \Wa_{\scriptscriptstyle R}^\dagger \left( (T^{(l,j)}_N)_{\scriptscriptstyle LR}^\dagger \widetilde{\mb{\helB}}_{\scriptscriptstyle L}^{l,\dagger} + (T^{(l,j)}_N)_{\scriptscriptstyle RR}^\dagger \widetilde{\mb{\helB}}_{\scriptscriptstyle R}^{l,\dagger} \right)\widetilde{\mb{\helB}}_{\scriptscriptstyle R}^j \Wa_{\scriptscriptstyle R},
	\end{multline}
	which, after orientation-averaging, yields $C^\text{(R)}_0$. Finally, the term $D^\text{(R)}$ is derived as follows:
\begin{widetext}
	\begin{align}
		D^\text{(R)} & = \sum_{\substack{i,j,k,l=1\\\mathclap{\text{pairwise distinct}}}}^{N} [0, \Wa_{\scriptscriptstyle R}^\dagger]
			\begin{bmatrix}
				J^{(k,l)}_{\scriptscriptstyle LL} & {0} \\  {0} & J_{\scriptscriptstyle RR}^{(k,l)}
			\end{bmatrix}
			\begin{bmatrix}
				(T^{(j,l)}_N)_{\scriptscriptstyle LL}^\dagger & (T^{(j,l)}_N)_{\scriptscriptstyle RL}^\dagger\\ (T^{(j,l)}_N)_{\scriptscriptstyle LR}^\dagger & (T^{(j,l)}_N)_{\scriptscriptstyle RR}^\dagger
			\end{bmatrix}
			\begin{bmatrix}
				\widetilde{\mb{\helB}}_{\scriptscriptstyle L}^{j,\dagger} \\ \widetilde{\mb{\helB}}_{\scriptscriptstyle R}^{j,\dagger}
			\end{bmatrix}
			[\widetilde{\mb{\helB}}_{\scriptscriptstyle L}^i, -\widetilde{\mb{\helB}}_{\scriptscriptstyle R}^i]
			\begin{bmatrix}
				(T^{(i,k)}_N)_{\scriptscriptstyle LL} & (T^{(i,k)}_N)_{\scriptscriptstyle LR}\\ (T^{(i,k)}_N)_{\scriptscriptstyle RL} & (T^{(i,k)}_N)_{\scriptscriptstyle RR}
			\end{bmatrix}
			\begin{bmatrix}
				0\\\Wa_{\scriptscriptstyle R}
			\end{bmatrix},
		\\
		&=\sum_{\substack{i,j,k,l=1\\\mathclap{\text{pairwise distinct}}}}^{N} \Wa_{\scriptscriptstyle R}^\dagger J_{\scriptscriptstyle RR}^{(k,l)}\left((T^{(j,l)}_N)_{\scriptscriptstyle LR}^\dagger \widetilde{\mb{\helB}}_{\scriptscriptstyle L}^{j,\dagger} + (T^{(j,l)}_N)_{\scriptscriptstyle RR}^\dagger\widetilde{\mb{\helB}}_{\scriptscriptstyle R}^{j,\dagger}\right) \left(\widetilde{\mb{\helB}}_{\scriptscriptstyle L}^i (T^{(i,k)}_N)_{\scriptscriptstyle LR} - \widetilde{\mb{\helB}}_{\scriptscriptstyle R}^i (T^{(i,k)}_N)_{\scriptscriptstyle RR}\right)\Wa_{\scriptscriptstyle R}.
	\end{align}
\end{widetext}
As above, the orientation average of this expression will result in $D^\text{(R)}_0$. The proof for left-handed incident fields works the same way with appropriate index changes.
\end{document}